\documentclass[tradiabstract]{aa}  

\usepackage{graphicx}
\usepackage{txfonts}
\usepackage{longtable}
\usepackage{lscape}
\usepackage{wasysym}

\newcommand{\G}{\mathrm{G}}
\newcommand{\dd}{\mathrm{d}}

\newcommand{\dm}{\dot{m}}
\newcommand{\Mjup}{$\mathrm{M_{\jupiter}}$}

\newcommand{\Msun}{$\mathrm{M_\odot}$}
\newcommand{\Rjup}{$\mathrm{R_{\jupiter}}$}

\newcommand{\Rsun}{$\mathrm{R_\odot}$}

\newcommand{\Lbol}{L_\mathrm{bol}}

\newcommand{\AU}{{\sc au}}
\newcommand{\etal}{et al.}
\newcommand{\gs}{g~s$^{-1}$}

\newcommand{\Rosat}{\emph{Rosat}}
\newcommand{\Xmm}{\emph{XMM-Newton}}
\newcommand{\xeuv}{{\mathrm{X/EUV}}}

\begin{document}

   \title{Mass-loss rates for transiting exoplanets}
   \author{D.~Ehrenreich\inst{1} \& J.-M.~D\'esert\inst{2}}

	\authorrunning{Ehrenreich \& D\'esert}

   \offprints{D.~Ehrenreich}

   \institute{
    	Institut de plan\'etologie et d'astrophysique de Grenoble (IPAG), Universit\'e Joseph Fourier-Grenoble~1, CNRS (UMR~5274), BP~53 38041 Grenoble CEDEX~9, France, \email{david.ehrenreich@obs.ujf-grenoble.fr}
     \and
     Harvard-Smithsonian Center for Astrophysics, 60 Garden street, Cambridge, Massachusetts 02138, USA, \email{jdesert@cfa.harvard.edu}
   }

   \date{}
 
  \abstract{Exoplanets at small orbital distances from their host stars are submitted to intense levels of energetic radiations, X-rays and extreme ultraviolet (EUV). Depending on the masses and densities of the planets and on the atmospheric heating efficiencies, the stellar energetic inputs can lead to atmospheric mass loss. These evaporation processes are observable in the ultraviolet during planetary transits. The aim of the present work is to quantify the mass-loss rates ($\dm$), heating efficiencies ($\eta$), and lifetimes for the whole sample of transiting exoplanets, now including hot jupiters, hot neptunes, and hot super-earths. The mass-loss rates and lifetimes are estimated from an ``energy diagram'' for exoplanets, which compares the planet gravitational potential energy to the stellar X/EUV energy deposited in the atmosphere. We estimate the mass-loss rates of all detected transiting planets to be within $10^6$ to $10^{13}$~\gs\ for various conservative assumptions. High heating efficiencies would imply that hot exoplanets such the gas giants WASP-12b and WASP-17b could be completely evaporated within 1~Gyr. We further show that the heating efficiency can be constrained when $\dm$ is inferred from observations and the stellar X/EUV luminosity is known. This leads us to suggest that $\eta \sim 100\%$ in the atmosphere of the hot jupiter HD~209458b, while it could be lower for HD~189733b. Simultaneous observations of transits in the ultraviolet and X-rays are necessary to further constrain the exospheric properties of exoplanets.}


   \maketitle
%

\section{Introduction}
\label{sec:intro}
More than $500$ extrasolar planets have been detected so far. A large fraction ($\sim30\%$) of them lie at close distances from their host stars, below $0.1$~\AU. Early on, the fact that most detected planets were so close to their stars \emph{and} massive enough to be most certainly gaseous, deeply questionned theoreticians about their origins and their fates. ``Hot jupiters'' cannot form at their observed locations, but have to migrate inward (Lin \etal\ 1996). Meanwhile, why did these planets stop migrating at such distances is still an open question. In contrast, one could wonder what happened to those giant planets that migrated further in. Of particular interest, is the question of the atmospheric stability at extreme levels of irradiation. How fast can giant planets be despoiled of their atmospheres? Couldn't lighter, Neptune- or Earth-mass planets form or migrate as well into environments that would questioned their atmospheric stability? Can such ``evaporation'' of exoplanets explain the apparent desert of planets below $\sim0.01$~\AU? In fact, the closest planets should also be the easiest to detect because the transit frequency and probability increase inversely to the semi-major axis.
 
Observations of exoplanet transits hold the keys to these questions. For the $>100$ known transiting planets, it is possible to measure the planet-to-star radius ratio and, for the few planets observed in the ultraviolet, constrain the size and mass-loss rate of their evaporating upper atmospheres. Transit observations of hot jupiters in the stellar Lyman-$\alpha$ emission of neutral hydrogen (H\,{\sc i}~Ly$\alpha$ at 1\,215.67~\AA) yield typical estimations of the mass-loss rates of $\dm\sim10^{10}$ to $10^{11}$~g~s$^{-1}$ for the planets \object{HD~209458b} (Vidal-Madjar \etal\ 2003, 2004, 2008; Ehrenreich \etal\ 2008; Linsky \etal\ 2010), \object{HD~189733b} (Lecavelier des Etangs \etal\ 2010), and \object{WASP-12b} (Fossati \etal\ 2010). Such values are predicted by numerous theoretical works and imply that the evolution of known hot jupiters is not significantly impacted by atmospheric evaporation (Lammer \etal\ 2003; Lecavelier des Etangs \etal\ 2004; Baraffe \etal\ 2004, 2005, 2006; Yelle 2004, 2006; Jaritz \etal\ 2005; Tian \etal\ 2005; Garc\'\i a-Mu\~noz 2007; Holmstr\"om \etal\ 2008; Stone \& Proga 2009; Murray-Clay \etal\ 2009). 

Lecavelier des Etangs (2007) compares the stellar energy received by the upper atmospheres to the gravitational energies of the planets in a so-called ``energy diagram''. This diagram allows estimations of the mass-loss rates and lifetimes for the population of close-in planets. This author uses the sample of exoplanets known as of June 2006, including ten transiting planets.  Recently, Davis \& Wheatley (2009) have provided an updated energy diagram containing 36 transiting hot jupiters, while Lammer \etal\ (2009) have determined the mass loss limit for 57 transiting planets. These studies pointed out that there should exist an evaporation-forbidden region into which a lost population of planets have been significantly eroded by this process, possibly leaving evaporation remnants, coined ``chthonian planets'' (Lecavelier des Etangs \etal\ 2004). 

In this article, we aim at providing comparative estimations of the mass-loss rates and lifetimes of the whole sample of transiting planets currently known in October 2010 (105 planets). As the sample of transiting planets steadily increases in time, the explored ranges of mass and density expand to new kinds of exoplanets: hot neptunes and hot super-earths. The existence of such close-in low-mass and high-density planets is a test for atmospheric evaporation theories. Moreover, predictions from the energy diagram enable to estimate the observable transit signature of evaporating planets (e.g., Ehrenreich \etal\ 2011).

The method to derive the energy diagram is detailed in Sect.~\ref{sec:diagram}. Afore-mentioned theoretical studies point out the role of X and extreme ultraviolet (EUV) stellar radiations as the main source of exospheric heating leading to mass loss. As recent X/EUV surveys were dedicated to exoplanet host stars (Kashyap \etal\ 2008; Sanz-Forcada \etal\ 2010; Poppenhaeger \etal\ 2010), we examine the impact of the constraints brought by these studies on the energy diagram (Sect.~\ref{sec:xflux}). In a few cases, measurements of the stellar X flux \emph{and} estimations of the mass-loss rate from observations of ultraviolet transits make it possible to constrain the mechanisms of atmospheric evaporation in terms of heating efficiency. This is discussed in Sect.~\ref{sec:eta}. Finally, the estimation of the atmospheric mass-loss rates raises questions about the atmospheric stability and the lifetime of evaporating planets, which is discussed in Sect.~\ref{sec:lifetime}.

\section{Updated energy diagram}
\label{sec:diagram}

The energy diagram for extrasolar planets measures the gravitational potential energy per mass unit of a planet, as a function of the X and EUV irradiations received by the planet. Next, we recall the main steps of the calculations further detailed in Lecavelier des Etangs (2007). All the properties of the planets and host stars used for these calculations are extracted from the \emph{Extrasolar Planets Encyclop\ae dia} (Schneider 2010) and are given in Table~\ref{tab:planets}.

\addtocounter{table}{1}


The framework is that of an energy-limited escape, i.e., an atmospheric mass unit can escape when the gravitational potential ($\dd E'_p$) is filled by the stellar X/EUV energetic input ($\dd E_\mathrm{X/EUV}$) into the planet upper atmosphere,
\begin{equation}
\label{eq:balance}
-\dd E'_p = \eta \dd E_\mathrm{X/EUV}.
\end{equation}

The key factor $\eta$ is the heating efficiency in the exoplanet thermosphere that encapsulates most of the physics of the problem. This factor value is high when most of the incoming stellar energy is effectively used to escape the atmosphere. It depends mainly on the fraction of energetic photons absorbed by atmospheric atoms, how deep in the atmosphere the absorption takes place, and how thick the absorbing region is. 

Lecavelier des Etangs (2007) consider the extreme case $\eta = 1$, where all the stellar flux is used to escape the atmosphere. Considering (without quantifying) energetic losses due to thermal emission by atmospheric hydrogen, Tian \etal\ (2005) model the atmospheric escape process of the hot jupiter HD~209458b using $\eta=0.15$. This value was initially chosen by Watson, Donahue \& Walker (1981) in their pioneering work about Earth's atmospheric escape. A similar value ($\eta=0.1$) was also employed by Valencia \etal\ (2010), who estimate the atmospheric loss for Corot-7b. In fact, the value of $\eta$ for the strongly irradiated atmospheres of extrasolar planets remains largely unconstrained (though see Cecchi-Pestellini \etal\ 2009 and Sect.~\ref{sec:eta}). In the following, we treat $\eta$ as a free parameter with a possible value between 0.01 and 1, until we detail in Sect.~\ref{sec:eta} how it could be possible to obtain observing constraints on this parameter. 

\begin{figure}
\resizebox{\hsize}{!}{\includegraphics{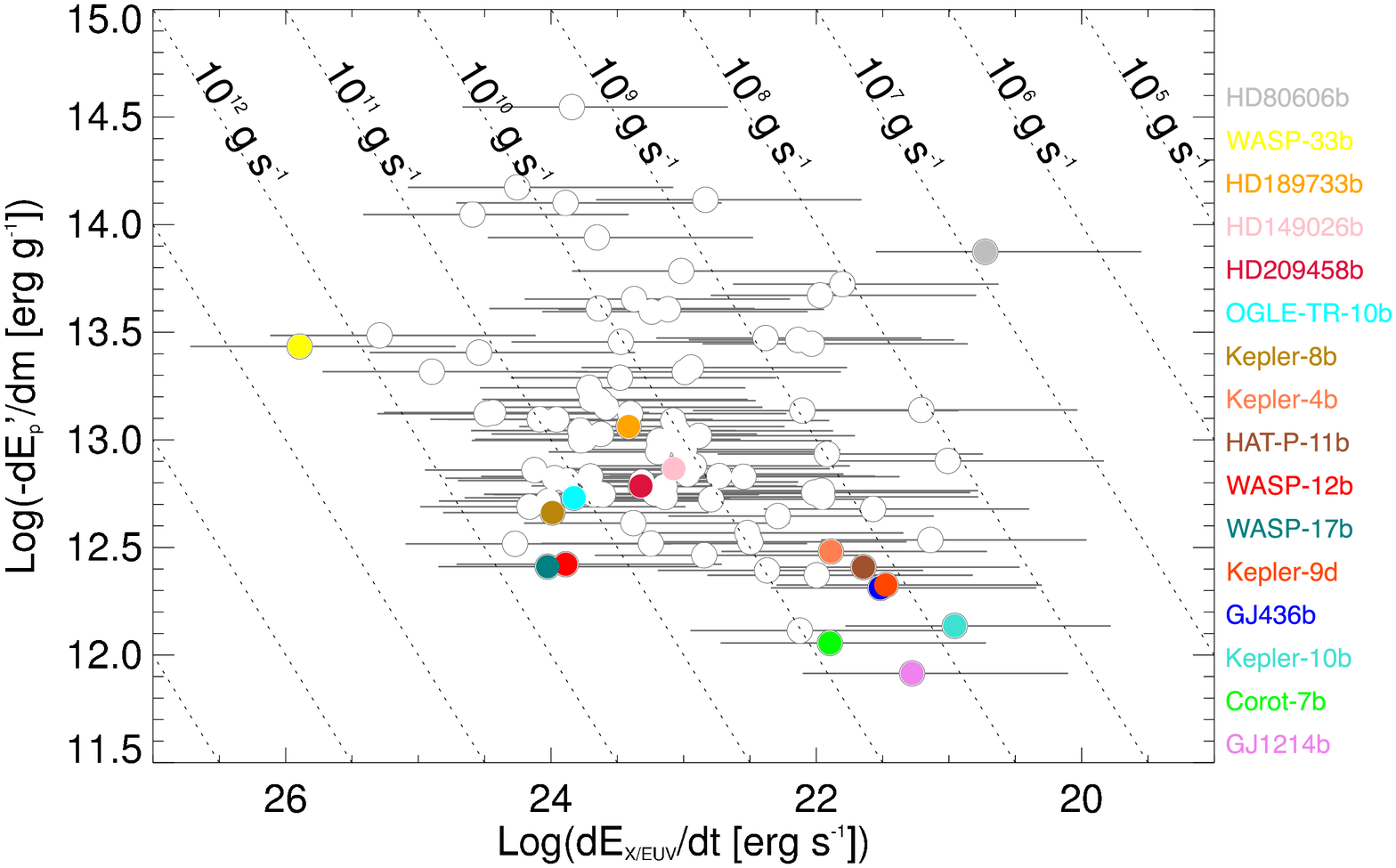}}
\caption{The energy diagram for 98 transiting planets: the energy needed to escape a unit of mass of the planet atmosphere versus the X/EUV flux reaching the top of the atmosphere per unit of time. The dotted lines indicates constant mass loss rates of $\dot{m} = 10^{14}$ to $10^{5}$~g~s$^{-1}$ (from left to right). Data points are calculated with $\eta=15\%$. Variations of $\eta$ between 1\% and 100\% are represented with horizontal gray error bars. Coloured points and labels indicate transiting planets of particular interest which are discussed in the text.\label{fig:EnergyDiagramXcal}}
\end{figure}

We present the updated energy diagram in Fig.~\ref{fig:EnergyDiagramXcal} with 98 transiting planets. The vertical axis of the energy diagram measures the potential energy required to escape the planetary atmosphere. The potential energy per atmospheric mass unit $\dd E_p'/\dd m$ of a planet of radius $R_p$ and mass $M_p$ includes the contribution of tidal forces and is calculated according to Eq.~(8) in Lecavelier des Etangs (2007).

The horizontal axis of the energy diagram measures the amount of stellar energy available on top of the planetary atmosphere. Cecchi-Pestellini \etal\ (2009) have studied how EUV photons (30--912~\AA) and X-rays (1--30~\AA) can heat the upper and lower atmospheric layers, respectively. This heating makes the atoms and ions of the upper atmosphere escaping the gravitational well of the planet. Lecavelier des Etangs (2007) estimates the amount of X/EUV energy received per unit of time as
\begin{equation} 
\label{eq:dEeuv}
\frac{\dd E_\xeuv}{\dd t} = \frac{1}{4} R_p^2 a_p^{-2} L_\xeuv.
\end{equation}
The X/EUV luminosity $L_\xeuv$ (1--912~\AA) is generally not known for host stars of transiting planets (but see Sect.~\ref{sec:xflux}). Lecavelier des Etangs (2007) uses the Wood \etal\ (1994) correlation between the stellar rotation velocity $v_\mathrm{rot}$ and the EUV flux $F_\mathrm{EUV}$ measured in the S2 bandpass of the \emph{Rosat} satellite (110--195~\AA) as a proxy to estimate the EUV flux of the 11 transited stars known in 2007,
\begin{equation}
\label{eq:vrot}
F_\mathrm{EUV}(1~\textrm{\AU}) =
4.6\left( \frac{v_\mathrm{rot}}{2.0~\textrm{km
      s}^{-1}}\right)^{1.4}~{\rm erg~cm^{-2}~s^{-1}}.
\end{equation}
The luminosity and flux at 1~\AU\ are related through $L_\mathrm{EUV} = 4\pi(\textrm{1~\AU})^2 F_\mathrm{EUV}(\textrm{1~\AU})$. We caution here that such an activity-rotation relation is only valid for stars with a convective zone. We therefore consider this estimation to be doubtful for stars earlier than $\sim$F5 (about 11\% of transited stars). In particular, the position of WASP-33b (A5 star) in Fig.~\ref{fig:EnergyDiagramXcal} should be regarded with caution.

\begin{figure}
\resizebox{\hsize}{!}{\includegraphics{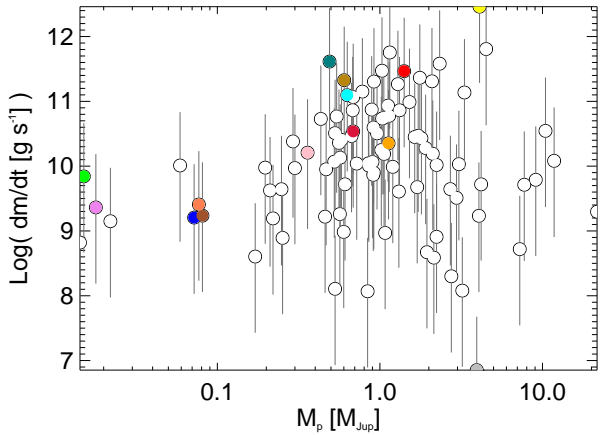}}
\caption{Mass-loss rate ($\eta = 0.15$) as a function of the planet mass for 98 transiting planets.  The vertical error bars represent a variation of $\eta$ from 0.01 (lower values) to 1 (higher values). Planets of interest are coloured accordingly to Fig.~\ref{fig:EnergyDiagramXcal}\label{fig:MassLossRateMass}}
\end{figure}

The measured $v_\mathrm{rot} \sin i_\star$ have been retrieved from the literature for 98 stars in Table~\ref{tab:planets}. We also assumed that the sine of the stellar inclination $i_\star$ is $\sim 1$ for transiting systems. The data points in Fig.~\ref{fig:EnergyDiagramXcal} are calculated assuming $\eta = 0.15$. The value of this factor dominates the uncertainties on such calculations. Therefore, the positions of planets in the diagram are shown with conservative error bars representing possible $\eta$ values in the range $[0.01,1]$. Curves of constant mass-loss rates $\dm$ are overplotted on Fig.~\ref{fig:EnergyDiagramXcal}. All transiting planets in the diagram have $\dm$ values between $10^{6}$~g~s$^{-1}$ and $10^{13}$~g~s$^{-1}$, assuming all possible values of $\eta$. 

Two `outliers', WASP-33b and HD~80606b, appear in the diagram with the highest and lowest mass-loss rates, respectively. Meanwhile, the mass-loss rate of WASP-33b is probably overestimated for the afore-mentioned reason. The mass-loss rate of the extremely eccentric ($e=0.93$) planet HD~80606b is likely underestimated, at least when the planet is at periastron, since the mass loss rates are calculated with the semi-major axis of transiting planets instead of the periastron distances.

The iso-mass-loss rates are represented by linear curves in Fig.~\ref{fig:EnergyDiagramXcal}. The mass-loss rate is also plotted as a function of the planet's mass in Fig.~\ref{fig:MassLossRateMass}. Both Figs.~\ref{fig:EnergyDiagramXcal} and~\ref{fig:MassLossRateMass} show that maximum mass loss rates ($10.5 \la \log \dm \la 12$) are reached with $\eta=0.15$ for planets with masses between $\sim0.4$ and $\sim4$~\Mjup. In Fig.~\ref{fig:MassLossRateMass}, there is a dearth of low-mass planets with high $\dm$, which is expected if low-mass planets are despoiled of their atmosphere at high $\dm$. The remaining cores (or chthonian planets) would be more difficult to detect in transit. Assuming a different value for $\eta$ would simply scale up or down the resulting mass loss rates, if one assumes a similar value of $\eta$ for all planets. See Sect.~\ref{sec:eta} for a discussion about this aspect.

\section{Measurements of X/EUV fluxes}
\label{sec:xflux}

To take into account the highest-energy photons able to heat the upper atmospheres of the transiting planets, it is necessary to estimate the X/EUV fluxes of planet-hosting stars. Observational and archival surveys have been started for this purpose. Kashyap \etal\ (2008) carried out a survey of X-ray emission combining archival and targeted data from \emph{Asca}, \emph{Exosat}, \emph{Einstein}, \emph{Rosat}, \emph{XMM-Newton}, and \emph{Chandra} missions. These authors present X-ray flux values or upper limits for 235 planet-hosting stars. These values cover a pass band of 0.1--4.5~keV (3--124~\AA). There are 35 transiting systems among their sample, among which only 5 have a detected X-ray emission (GJ~436, HD~209458, HD~189733, OGLE-TR-10, and SWEEPS-11).

Sanz-Forcada \etal\ (2010) study a sample of 65 planetary systems with \emph{XMM-Newton} and \emph{Chandra} archival data. The data sets are the same as in Kashyap \etal\ (2008) but some estimations of stellar X luminosities are revised. According to Sanz-Forcada \etal\ (2010), these differences appear in some cases (noticeably for HD~209458) where the stellar proper motion is not taken into account, leading to suspicious source identification. Concerning transiting systems, Sanz-Forcada \etal\ (2010) report lower values than Kashyap \etal\ (2008) for HD~189733 (by 0.2~dex) and GJ~436 (by 1~dex). They give an upper limit for HD~209458, significantly lower (by 0.7~dex) than the values reported elsewhere.

Poppenhaeger \etal\ (2010) present a similar work as the above-quoted studies. They add new \emph{XMM-Newton} data points for stars previously devoided of X-ray measurements. Unfortunately, their new data set does not encompass additional transiting systems. They report similar X-ray luminosity values for HD~189733 and GJ~436, although their calculated uncertainties are larger than those estimated in Kashyap \etal\ (2008). 

For some active stars, the X-ray luminosity may vary outside the range allowed by the uncertainties  reported in these works. In fact, a detailed analysis of archival \emph{XMM-Newton} observations of the active star HD~189733 at two different epochs show that these uncertainties probably underestimate the intrinsic source variability (Pillitteri \etal\ 2010). Pillitteri \etal\ (2010) indeed find that the X-ray flux of this star has variated by 45\% between 2007 and 2009.
 
\begin{table*}
\caption{Multiple X-ray detection reports in the literature for stars hosting transiting planets\label{tab:multix}}
\begin{tabular}{lcccccc}
\hline
\hline
Star                && \multicolumn{4}{c}{$\log (L_X \mathrm{[erg~s^{-1}]})$}                         \\
\cline{3-7}
                    && H\"unsch \etal\ & Kashyap \etal\   & Sanz-Forcada \etal\ & Poppenhaeger \etal\ & Pillitteri \etal\ \\
                    && (1999)          & (2008)           & (2010)              & (2010)              & (2010)            \\
\hline     
HD~189733           && $28.44^a$       & $28.43\pm0.01^a$ & $28.18^b$           & $28.26\pm0.12^b$    & $27.72^b$--$28.29^b$  \\
HD~209458           &&                 & $27.02\pm0.20^b$ & $<26.12^b$          &                     &                   \\
GJ~436              && $26.85^a$       & $27.16\pm0.19^a$ & $25.96^b$           & $27.16\pm0.34^a$    &                   \\
\hline
\multicolumn{7}{l}{\parbox{10cm}{\textbf{Notes.} $^a$ \Rosat /PSPC. $^b$ \Xmm /EPIC.}}
\end{tabular}
\end{table*}

\begin{figure}
\resizebox{\hsize}{!}{\includegraphics{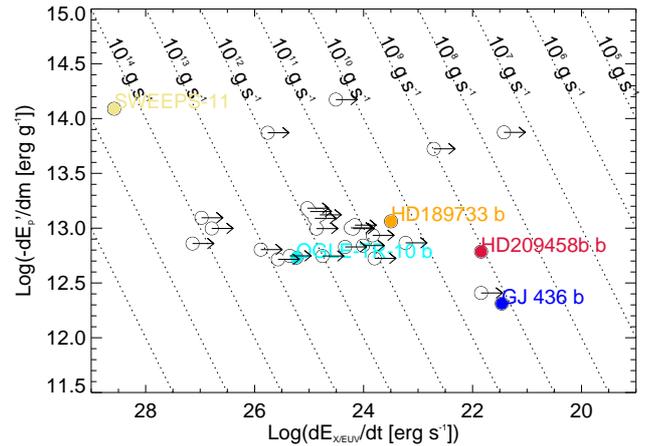}}
\caption{The energy diagram for 35 transiting planets with X-ray emission detections or upper limits (arrows) reported in the literature. Caption is the same as for Fig.~\ref{fig:EnergyDiagramXcal}. Data points are calculated with $\eta=15\%$. Planets of interest are outlined with different colours.\label{fig:EnergyDiagramXobs}}
\end{figure}

The values of $L_\mathrm{X/EUV}$ found in the literature have been injected in Eq.~(\ref{eq:dEeuv}) to obtain the energy diagram shown in Fig.~\ref{fig:EnergyDiagramXobs}. Among the five transiting systems with measured X-ray emissions (Kashyap \etal\ 2008), HD~189733b and GJ~436b stay at about the same location as in Fig.~\ref{fig:EnergyDiagramXcal} for $\eta=0.15$, with $\dm$ between $10^{10}$ and $10^{11}$~g~s$^{-1}$ for HD~189733b and $\sim10^9$~g~s$^{-1}$ for GJ~436b. In contrast, the mass loss rate of HD~209458b is about 1.5 order-of-magnitude lower than with the previous estimation. Noticeably, the $\dm$ value for OGLE-TR-10b dramatically increases by $\sim2$ orders of magnitude. Finally, the case of SWEEPS-11 is troublesome. There is no reported value for the rotational velocity of this star in the literature, so we were not able to include this object in Fig.~\ref{fig:EnergyDiagramXcal}. The X/EUV flux reported by Kashyap \etal\ (2008) for this star is more than 10 times higher than for HD~189733. SWEEPS-11 is a $20^\mathrm{th}$-magnitude star belonging to a crowded region in the Galactic bulge.\footnote{This comment could also apply to OGLE-TR-10.} The possibility that the detected emission does not come from this star is therefore not negligible.

Overall, there are multiple reports of the X-ray emissions of only three transiting systems in the literature. These are HD~189733, HD~209458, and GJ~436. The different values reported for these three systems are summarized in Table~\ref{tab:multix}.



\section{Constraints on the heating efficiency}
\label{sec:eta}

For the majority of transiting planets, the two largest uncertainties in the calculation of $\dm$ are the heating efficiency $\eta$ and the X/EUV luminosity. Cecchi-Pestellini \etal\ (2009) have studied the relative role of X/EUV radiation in the heating of the hydrogen-rich atmosphere of HD~209458b. In their model, the flux of stellar X/EUV photons incident upon the planetary atmosphere of solar-like composition photo-ionizes the gas producing a flux of high energy photo-electrons, which deposit their energy into the gas. They have calculated heating efficiencies as a function of photon energy, electron fraction in the gas, and vertical column density of the atmosphere. The results strongly depend on the coupled ionization and density structures as shown by Yelle (2004), Garc\'\i a-Mu\~noz (2007), and Murray-Clay \etal\ (2009). For a fractional electron concentration of 10\%, Cecchi-Pestellini \etal\ (2009) found heating efficiencies of $\sim0.9$, 0.75, and 0.7 for column densities $\ga10^{19}$, $10^{21}$, and $10^{22}$~cm$^{-2}$, respectively, and for photon energies of 50~eV ($\lambda=248$~\AA), 300~eV (41~\AA), and 1\,000~eV (12~\AA), respectively. These efficiencies decrease to 0.18, 0.04, and 0.02 when the electron fraction is $10^{-6}$. 

The work of Cecchi-Pestellini \etal\ (2009) shows that depending on the atmospheric properties, the heating efficiency can take nearly all possible values. Reversly, measuring the heating efficiency $\eta$ could provide some constraints on these atmospheric properties. In the few cases where transits have been observed in the UV, it is possible to constrain $\dm$. This can be done (i) numerically using a particle simulation with $\dm$ set as a free parameter to reproduce the absorption during transit caused by the cloud of escaping hydrogen atoms (Lecavelier des Etangs \etal\ 2010), or (ii) analytically by assuming a density profile --\,usually a power-law\,-- for the hydrodynamically escaping atmosphere (Linsky \etal\ 2010). If the X/EUV luminosity is known, UV transits can thus bring constraints on $\eta$. In fact, starting from Eq.~(\ref{eq:balance}), the heating efficiency can be expressed as the ratio of the two powers $L_\mathrm{X/EUV}$ and $L_{\dm}$ depending on the star and planet properties, respectively,
\begin{equation}
\eta = \frac{L_{\dm}}{L_\mathrm{X/EUV}}.  
\label{eq:eta}
\end{equation}
If tidal forces are neglected, the ``mass loss power'' $L_{\dm}$ has a simple expression,
\begin{equation}
\label{eq:Ldm}
L_{\dm} = (16\pi/3) \G \dm \rho_p a_p^2,
\end{equation} 
where $\rho_p = M_p/(4/3\pi R_p^3)$ is the mean density of the planet.

\begin{figure}
\resizebox{\hsize}{!}{\includegraphics{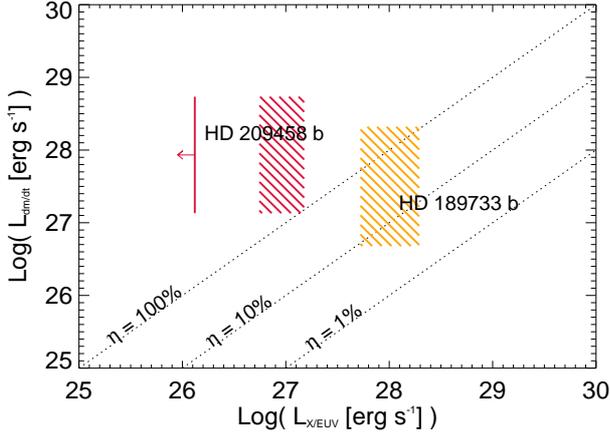}}
\caption{The power diagram for evaporating extrasolar planets. The ``mass-loss power'' $L_{\dm}$ (Eq.~\ref{eq:Ldm}) is plotted as a function of the X/EUV luminosity. Constant $\eta$ values are represented by the dotted lines. The hatched regions show the position of HD~189733b (orange) and HD~209458b (crimson). For HD~209458b, the right-pointing arrow show the upper limit on the X/EUV luminosity reported by Sanz-Forcada \etal\ (2010).\label{fig:powerdiag}}
\end{figure}

In the following, all the calculations are performed taking tidal forces into account. To constrain the value of $\eta$, we introduce in Fig.~\ref{fig:powerdiag} a ``power diagram'' for extrasolar planets. Only two transiting planets have both observational constraints on $\dm$ and $L_\mathrm{X/EUV}$: HD~209458b and HD~189733b. For HD~209458b, we have considered the value of $10^{10}$~\gs\ for the mass-loss rate given by Vidal-Madjar \etal\ (2003) as a lower limit. This value results from the $(15\pm4)\%$ absorption detected over $\sim1/3$ of the \ion{H}{i}~Lyman~$\alpha$ line. It is a lower limit because there is no way to measure the absorption over the central part of the line hidden by the interstellar medium absorption. Higher $\dm$ values, $(8$--$40)\times10^{10}$~\gs, have recently been reported by Linsky \etal\ (2010). As discussed below, these authors have inferred hydrogen mass-loss rates from a $(8\pm1)\%$ absorption in the emission line of singly ionized carbon (\ion{C}{ii}) during the transit of the planet.

Lecavelier des Etangs \etal\ (2010) have estimated the mass-loss rate of HD~189733b from a \ion{H}{i} Ly$\alpha$  transit light curve, for different values of the X/EUV flux. For a best-fit value $F_\mathrm{X/EUV}$ of 20 times the solar value, they report a 1-$\sigma$ range of $3.2\times10^9<\dm<1.4\times10^{11}$~\gs.

The X-ray luminosities are taken from Kashyap \etal\ (2008) for HD~209458 and Pillitteri \etal\ (2010) for HD~189733. A weakly efficient evaporation, with $\eta \sim 1\%$, seems to excluded from the observations of HD~189733. In contrast, only values of $\eta > 1$ are compatible with the observations of HD~209458b, which is physically not plausible. In fact, it is not clear what possible energy source could significantly cumulate with the stellar radiation to drive such a ``super-efficient'' mass loss. The situation is even more troublesome for HD~209458b if one considers the upper limit on $L_\mathrm{X/EUV}$ given by Sanz-Forcada \etal\ (2010). 

The uncertainties on the measured values of $L_\mathrm{X/EUV}$ and $\dm$ are a first possible explanation to the apparent ``super-efficiency'' of HD~209458b's mass loss. In fact, a large part of the EUV spectrum (roughly from 200 to 912~\AA) does not contribute to the ``X/EUV'' flux measured with \Rosat\ or \Xmm. If HD~209458 is as luminous between 200 and 912~\AA\ as it is below 200~\AA, then the position of the planet in the power diagram (Fig.~\ref{fig:powerdiag}) would become compatible with a very efficient ($\eta\sim100\%$) mass loss. However, in the case of the lower luminosity estimation of Sanz-Forcada \etal\ (2010), the star would have to be orders of magnitude more luminous in the EUV above 200~\AA\ than below. 

Another possible explanation for the super-efficiency of HD~209458b's mass loss may be found in the way the mass-loss rate was estimated by Linsky \etal\ (2010), which strongly depends on the assumed density profile of the exosphere, up to the Roche lobe (A.~Lecavelier des Etangs, private communication).

It is also possible that $L_\mathrm{X/EUV}$ was larger during the observations of Linsky \etal\ (2010) than in 2003. An increase by 1 or 2 order(s) of magnitude, from the value of Kashyap \etal\ (2008) or Sanz-Forcada \etal\ (2010), respectively, is needed to have $\eta$ values below 1, within the error bars. Such a large X luminosity variation might not be expected for a solar-type star like HD~209458. Meanwhile, the only way to solve this question is to obtain simultaneous measurements of the X-ray flux and the escape rate. This should be possible with current space instrumentation.

Assuming the X/EUV luminosity from Kashyap \etal\ (2008) and the escape rate $\dm$ between $10^{10}$ and $4\times10^{11}$~\gs, it is possible to infer exospheric properties out of the power diagram. In fact, while the $\dm$ estimation of Vidal-Madjar \etal\ (2003), marginally compatible with $\eta = 1$, is based on the observation of an absorption in the \ion{H}{i} stellar emission, the higher estimation of Linsky \etal\ (2010) is derived from the observation of an absorption in the singly ionized carbon (\ion{C}{ii}) stellar emission. For this line, these authors measured an $(8\pm1)\%$ absorption, yielding $\dm(\textrm{\ion{C}{ii}}) = 2.1\times10^7$~g~s$^{-1}$. To go from this value to a total (hydrogen) mass-loss rate $\dm \approx 8\times10^{10}$~g~s$^{-1}$, they rely on the assumptions that (i) all the carbon is ionized (i.e., $\rm[\ion{C}{ii}/C] = 1$) and (ii) that the carbon abundance is solar ($\rm [C/H]_\odot = 2.7\times10^{-4}$; Asplund \etal\ 2009) in the upper atmosphere of HD~209458b. The total (hydrogen) mass-loss rate is then $\dm \equiv \dm(\textrm{\ion{H}{i}}) = \dm(\textrm{\ion{C}{ii}})/\textrm{[C/H]}$. 

Therefore, an apparent super-efficient evaporation process with $\eta > 1$, as shown in Fig.~\ref{fig:powerdiag} could mean that $\rm [C/H]$ is underestimated. The $\rm [C/H]$ ratio would have to be increased by at least an order of magnitude with respect to $\rm [C/H]_\odot$ to give a $\dm$ value compatible with $\eta = 1$, assuming $L_\mathrm{X/EUV}$ values from Kashyap \etal\ (2008). On the other hand, if $\log L_\mathrm{X/EUV} < 26.12$ (Sanz-Forcada \etal\ 2010), then the upper atmosphere should be extremely carbon-rich with $\rm [C/H] / [C/H]{_\odot} > 100$. An underestimated $\rm [C/H]$ could also imply that the carbon ionization fraction is unity, as predicted by models of HD~209458b's atmosphere (Garc\'\i a-Mu\~noz 2007). 

Nevertheless, we caution that the measurements of $\dm$ and $L_\mathrm{X/EUV}$ for transiting planets may not be accurate enough at the present time to allow firm conclusions to be drawn out of the power diagram. Meanwhile, this shows how it could be possible to constrain the exospheric composition when precise measurements are available.

\section{Stability of evaporating atmospheres}
\label{sec:lifetime}

The lifetime of evaporating gaseous planets can be estimated as the time needed to exhaust the available reservoir of gas for a given mass-loss rate. An approximation of this life time can be obtained in an ``integrated'' version of the energy diagram shown in Fig.~\ref{fig:EnergyDiagramXcal}. We use the total potential energy $E'_p$ of a planet including tidal forces, defined by Eq.~(10) in Lecavelier des Etangs (2007).

The mean energy received $\langle\dd E_\mathrm{X/EUV}/\dd t\rangle$ is calculated by integrating Eq.~(\ref{eq:dEeuv}) over a given time interval. Doing so, one should account for the evolution of $L_\xeuv(t)$ in time. In fact, the high-energy radiation output from a star is stronger when the star is younger and varies with time as a power-law $L_\xeuv(t > \tau_\mathrm{sat}) \propto t^{-\alpha}$ (e.g., Penz \etal\ 2008), with $\alpha$ close to 1. Stars younger than $\tau_\mathrm{sat}$ are in the ``saturation regime'' where the evolution of $L_\xeuv(t)$ is almost flat. As discussed by Davis \& Wheatley (2009), the duration $\tau_\mathrm{sat}$ of the saturation period depends on the spectral type (see, e.g., Reiners \etal\ for M dwarfs). Here, we consider the equations used by Sanz-Forcada \etal\ (2010)
\begin{eqnarray}
\label{eq:Lsat}L_\xeuv & = & 6.3\times10^{-4} \Lbol \qquad (t < \tau_\mathrm{sat}), \\
\label{eq:Lxuv}L_\xeuv & = & 1.89 \times10^{28} t^{-1.55} \quad (t > \tau_\mathrm{sat}), \\
\label{eq:tsat}\tau_\mathrm{sat} & = & 2.03\times10^{20} \Lbol^{-0.65}.
\end{eqnarray}
We use this set of equations to estimate $\langle L_\xeuv \rangle$, the mean amount of X/EUV irradiation received during the life time of our sample stars per billion year,
\begin{equation}
\label{eq:meanLxeuv}
\langle L_\xeuv \rangle = \frac{1}{\tau_\star} \left( \int_{0}^{\tau_\mathrm{sat}} L_\xeuv(t)\dd t + \int_{\tau_\mathrm{sat}}^{\tau_\star} L_\xeuv(t) \dd t \right).
\end{equation}
After injecting Eqs.~(\ref{eq:Lsat}) to~(\ref{eq:tsat}) into Eq.~(\ref{eq:meanLxeuv}), we obtain
\begin{eqnarray*}
\langle L_\xeuv \rangle & = & \frac{1}{\tau_\star} \left[ 1.28\times10^{17}\Lbol^{0.35} + 2.33\times10^{17}\Lbol^{0.3575} \right. \\
                        & - & \left. 3.44\times10^{28}\tau_\star^{-0.55} \right],
\end{eqnarray*}
where $\Lbol$ and $\tau_\star$ are the present-day bolometric luminosity in erg~s$^{-1}$ and age of the star in Gyr, respectively. We calculate $\Lbol$ for our sample stars with Stefan-Boltzmann's law, $\Lbol = 4\pi \sigma T_\mathrm{eff}^4 R_\star^2$, where the effective temperatures and stellar radii are taken from the literature and listed in Table~\ref{tab:planets}. For M, K, G, F, and A stars, we assumed typical stellar ages of 10, 7, 5, 3, and 1~Gyr, respectively.

We want to know by how much the X/EUV irradiation is underestimated when the time evolution of $L_\xeuv$ is ignored. For this purpose, we introduce the factor $\gamma$ (see also Lecavelier des Etangs 2007),
\begin{equation}
\gamma = \frac{\langle L_\xeuv \rangle}{L_\xeuv},
\end{equation}
where $L_\xeuv$ at the denominator is calculated using Eq.~(\ref{eq:Lxuv}) with $t=\tau_\star$. The values of $\gamma$ are listed in Table~\ref{tab:planets}. We found a mean value of $\gamma = 38\pm9$ ($1\sigma$). We apply this $\gamma$ correction to the values of $\langle\dd E_\mathrm{X/EUV}/\dd t\rangle$ calculated by integrating Eq.~(\ref{eq:dEeuv}).

This parameterization neglects the fact that the heating efficiency might be notably smaller at early times when the stellar luminosity is higher. Under high X/EUV illumination, Murray-Clay \etal\ (2009) have indeed suggested that the mass-loss regime should depart from an energy-limited escape ($\dm \propto F_\mathrm{X/EUV}^{0.9}$) and instead become ``radiation/recombination''-limited ($\dm \propto F_\mathrm{X/EUV}^{0.6}$). This effect would increase the lifetimes calculated here, where we assume that the heating efficiency remains constant over time.

\begin{figure}
\resizebox{\hsize}{!}{\includegraphics{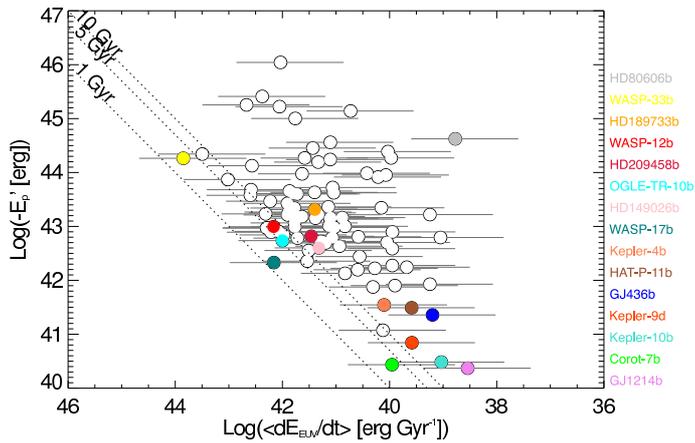}}
\caption{The ``integrated'' energy diagram for 98 transiting planets: the energy needed to escape the whole planet mass versus the X/EUV flux received per billion year. The dotted lines represent constant lifetimes of 1, 5, and 10~Gyr (from bottom to top). Data points are calculated with $\eta = 0.15$. Variations of $\eta$ between 0.01 and 1 are represented as horizontal gray error bars. \label{fig:TotalEnergyDiagram}} 
\end{figure}

These integrated energies are calculated for our sample of transiting planets and the result is shown in Fig.~\ref{fig:TotalEnergyDiagram} for different values of $\eta$. The ratio between $-E'_p$ (in erg) and $\langle \dd E_\xeuv / \dd t \rangle$ (in erg~Gyr$^{-1}$) is equivalent to a lifetime. As can be seen in Fig.~\ref{fig:TotalEnergyDiagram}, only a few planets have lifetimes below 10~Gyr. Meanwhile, this figure also highlights that any firm conclusions have to rely on an assumed value of $\eta$. For instance, if $\eta$ is close to $1$ for WASP-17b, WASP-33b, or WASP-12b, then the atmospheres of these planets should be lost in less than 1~Gyr. Noteworthy, WASP-12b may well fill its Roche lobe fully enough that it is losing mass faster than the process considered here would predict (Li \etal\ 2010).

An interesting case is that of Corot-7b. If this planet were made of gas, its whole mass would be evaporated within $\la 5$~Gyr for $\eta \ga 0.15$. Considering that Corot-7b is probably made of rock (L\'eger \etal\ 2009, 2011; Queloz \etal\ 2009), this means that the atmosphere of this planet could have been completely evaporated. This would make Corot-7b the first evaporation remnant detected. Possible origins of Corot-7b are discussed by Jackson \etal\ (2010). These authors conclude that this planet could have indeed started with a gas envelope, which total mass should have been $< 2/3$~\Mjup. 

\section{Conclusion}
\label{sec:conclu}

We have calculated mass-loss rates for transiting exoplanets using an energy diagram. This approach assumes that mass loss occurs as a consequence of a massive stellar irradiation. Mass-loss rates for the complete sample of transiting planets are found between $10^6$ and $10^{13}$~\gs. This (large) range encompass all possible values of the atmospheric heating efficiency. 

We have shown that measurements of the heating efficiency can be constrained in a power diagram when the mass-loss rate, derived from observations of an evaporating atmospheres, and the X/EUV flux are both estimated. At the present time, it is only possible to spot two planets in this diagram. While the precision on heating efficiency is limited  by the estimation of the mass-loss rate, we have shown how it could be possible to infer exospheric properties, such as relative abundances, out of more precise measurements. In particular, simultaneous transit observations in UV and X-rays, feasible with present-time space instrumentation, will bring stronger constraints on the atmospheric heating efficiencies.

The mass-loss rates are used to determine the lifetimes of evaporating planets. We have found that few planets have lifetimes below 1~Gyr but this requires a highly efficient heating with $\eta \sim 1$. 

\acknowledgements
We thank the anonymous referee of this work. We are grateful to Alain Lecavelier des Etangs for numerous discussions and advices about this work. We also thank Alfred Vidal-Madjar for inspiration. D.E. is supported by the Centre National d'\'Etudes Spatiales (CNES).

\longtab{1}{
\begin{landscape}
\begin{longtable}{l*{15}{c}}
\caption{\label{tab:planets} Properties of known transiting planets and their host stars.}\\
\hline \hline
     & \multicolumn{4}{c}{Planet}      & & \multicolumn{7}{c}{Host star}                                                                                           &                                         &          \\
\cline{2-5} \cline{7-14}
Name & $M_p$   & $R_p$   & $P$ & $a_p$ & & Sp.T. & $M_\star$ & $R_\star$ & $V$   & $v_\mathrm{rot} \sin i_\star$ & $T_\mathrm{eff}$ & Ref.  & ${F_\mathrm{X/EUV}}$ & $\log \dm$                              & $\gamma$ \\
     & (\Mjup) & (\Rjup) & (d) & (\AU) & &       & (\Msun)   & (\Rsun)   & (mag) & (km s$^{-1}$)                 & (K)              &       & (1)                  & $\frac{\eta}{0.15}$(\gs)                &          \\
\hline
CoRoT-1 b & 1.03 & 1.49 & 1.50 & 0.025 &  &G0V & 0.95 & 1.11 & 13.6 & 5.20 &5950 &2 & &11.4 &42.6 \\
CoRoT-10 b & 2.75 & 0.97 & 13.2 & 0.105 &  &K1V & 0.89 & 0.79 & 15.2 & 2.00 &5075 &3 & &8.30 &31.7 \\
CoRoT-11 b & 2.33 & 1.43 & 2.99 & 0.043 &  &F6V & 1.27 & 1.37 & 12.9 & 40.0 &6440 &4 & &11.5 &41.8 \\
CoRoT-12 b & 0.91 & 1.44 & 2.82 & 0.040 &  &G2V & 1.07 & 1.11 & 15.5 & 1.00 &5675 &5 & &9.90 &39.9 \\
CoRoT-13 b & 1.30 & 0.88 & 4.03 & 0.051 &  &G0V & 1.09 & 1.01 & 15.0 & 4.00 &5945 &6 & &9.60 &39.7 \\
CoRoT-2 b & 3.31 & 1.46 & 1.74 & 0.028 &  &G7V & 0.97 & 0.90 & 12.5 & 11.5 &5625 &7 & &11.1 &33.6 \\
CoRoT-3 b & 21.6 & 1.01 & 4.25 & 0.057 &  &F3V & 1.37 & 1.56 & 13.3 & 17.0 &6740 &8 & &9.29 &49.2 \\
CoRoT-4 b & 0.72 & 1.19 & 9.20 & 0.090 &  &F0V & 1.10 & 1.15 & 13.7 & 6.40 &6190 &9 & &10.0 &34.6 \\
CoRoT-5 b & 0.46 & 1.38 & 4.03 & 0.049 &  &F9V & 1.00 & 1.18 & 14.0 & 1.00 &6100 &10 & &9.95 &34.6 \\
CoRoT-6 b & 2.96 & 1.16 & 8.88 & 0.085 &  &F5V & 1.05 & 1.02 & 13.9 & 7.50 &6090 &11 & &9.51 &30.9 \\
CoRoT-7 b & 0.01 & 0.15 & 0.85 & 0.017 &  &K0V & 0.93 & 0.87 & 11.7 & $<$3.50 &5275 &12 & &9.84 &36.1 \\
CoRoT-8 b & 0.22 & 0.57 & 6.21 & 0.063 &  &K1V & 0.88 & 0.77 & 14.8 & 2.00 &5080 &13 & &9.19 &31.1 \\
CoRoT-9 b & 0.84 & 1.05 & 95.2 & 0.407 &  &G3V & 0.99 & 0.94 & 13.7 & $<$3.50 &5625 &14 & &8.06 &34.6 \\
GJ 1214 b & 0.01 & 0.24 & 1.58 & 0.014 &  &M4.5 & 0.15 & 0.21 & 14.6 & $<$2.00 &3026 &15 & &9.36 &5.82 \\
GJ 436 b & 0.07 & 0.43 & 2.64 & 0.028 &  &M2.5 & 0.45 & 0.46 & 10.6 & $<$3.00 &3585 & 16 , 17  &1.17$\pm$0.52 &9.20 &15.2 \\
HAT-P-1 b & 0.52 & 1.22 & 4.46 & 0.055 &  &G0V & 1.13 & 1.11 & 10.4 & 2.20 &6047 &18 &$<$0.64 &10.0 &43.8 \\
HAT-P-11 b & 0.08 & 0.45 & 4.88 & 0.053 &  &K4 & 0.81 & 0.75 & 9.59 & 1.50 &4780 &19 &$<$0.64 &9.23 &27.8 \\
HAT-P-12 b & 0.21 & 0.95 & 3.21 & 0.038 &  &K5 & 0.73 & 0.70 & 12.8 & 0.50 &4650 &20 & &9.62 &25.3 \\
HAT-P-13 b & 0.85 & 1.28 & 2.91 & 0.042 &  &G4 & 1.22 & 1.56 & 10.6 & 1.76 &5653 &21 & &10.0 &50.8 \\
HAT-P-14 b & 2.23 & 1.15 & 4.62 & 0.060 &  &F & 1.38 & 1.46 & 9.98 & 8.40 &6600 &22 & &10.0 &45.6 \\
HAT-P-15 b & 1.94 & 1.07 & 10.8 & 0.096 &  &G5 & 1.01 & 1.08 & 12.1 & 2.00 &5568 &23 & &8.67 &37.8 \\
HAT-P-16 b & 4.19 & 1.28 & 2.77 & 0.041 &  &F8 & 1.21 & 1.23 & 10.8 & 3.50 &6158 &24 & &9.71 &36.2 \\
HAT-P-17 b & 0.53 & 1.01 & 10.3 & 0.088 &  &K & 0.85 & 0.83 & 10.5 & 0.30 &5246 &25 & &8.10 &34.7 \\
HAT-P-18 b & 0.19 & 0.99 & 5.50 & 0.055 &  &K & 0.77 & 0.74 & 12.7 & 1.50 &4750 &26 & &9.97 &27.5 \\
HAT-P-19 b & 0.29 & 1.13 & 4.00 & 0.046 &  &K & 0.84 & 0.82 & 12.9 & 2.10 &4875 &26 & &10.3 &30.6 \\
HD 147506 b & 9.09 & 1.15 & 5.63 & 0.068 &  &F8 & 1.36 & 1.64 & 8.71 & 19.8 &6290 &27 & &9.78 &46.1 \\
HAT-P-20 b & 7.24 & 0.86 & 2.87 & 0.036 &  &K3 & 0.75 & 0.69 & 11.3 & 2.10 &4595 &28 & &8.71 &24.7 \\
HAT-P-21 b & 4.06 & 1.02 & 4.12 & 0.049 &  &G3 & 0.94 & 1.10 & 11.6 & 3.50 &5588 &28 & &9.23 &38.7 \\
HAT-P-22 b & 2.14 & 1.08 & 3.21 & 0.041 &  &G5 & 0.91 & 1.04 & 9.73 & 0.50 &5302 &28 & &8.58 &34.2 \\
HAT-P-23 b & 2.09 & 1.36 & 1.21 & 0.023 &  &G0 & 1.13 & 1.20 & 12.4 & 8.10 &5905 &28 & &11.3 &44.7 \\
HAT-P-24 b & 0.68 & 1.24 & 3.35 & 0.046 &  &F8 & 1.19 & 1.31 & 11.8 & 10.0 &6373 &29 & &11.0 &40.0 \\
HAT-P-25 b & 0.56 & 1.19 & 3.65 & 0.046 &  &G5 & 1.01 & 0.95 & 13.1 & 0.50 &5500 &30 & &9.26 &34.0 \\
HAT-P-26 b & 0.05 & 0.56 & 4.23 & 0.047 &  &K1 & 0.81 & 0.78 & 11.7 & 1.80 &5079 &31 & &10.0 &31.6 \\
HAT-P-3 b & 0.59 & 0.89 & 2.89 & 0.038 &  &K & 0.93 & 0.82 & 11.8 & 0.50 &5185 &32 &$<$0.64 &8.98 &33.7 \\
HAT-P-4 b & 0.68 & 1.33 & 3.05 & 0.044 &  &F & 1.26 & 1.60 & 11.2 & 5.50 &5860 &33 &$<$0.64 &10.8 &40.8 \\
HAT-P-5 b & 1.06 & 1.26 & 2.78 & 0.040 &  &G & 1.16 & 1.16 & 12.0 & 2.60 &5960 &34 &$<$0.64 &10.1 &44.4 \\
HAT-P-6 b & 1.05 & 1.33 & 3.85 & 0.052 &  &F & 1.29 & 1.46 & 10.5 & 8.70 &6570 &35 &$<$0.64 &10.7 &45.2 \\
HAT-P-7 b & 1.80 & 1.42 & 2.20 & 0.037 &  &F6 & 1.47 & 1.84 & 10.5 & 3.80 &6350 &36 &$<$0.64 &10.4 &50.9 \\
HAT-P-8 b & 1.52 & 1.50 & 3.07 & 0.048 &  &F & 1.28 & 1.58 & 10.1 & 11.5 &6200 &37 & &10.9 &43.9 \\
HAT-P-9 b & 0.78 & 1.40 & 3.92 & 0.053 &  &F & 1.28 & 1.32 &  & 11.9 &6350 &38 & &11.1 &39.8 \\
HD 149026 b & 0.35 & 0.65 & 2.87 & 0.043 &  &G0 IV & 1.30 & 1.49 & 8.15 & 6.00 &6147 &39 &$<$1.15 &10.2 &55.8 \\
HD 17156 b & 3.21 & 1.02 & 21.2 & 0.162 &  &G0 & 1.24 & 1.44 & 8.17 & 2.60 &6079 &40 &$<$2.06 &8.07 &53.5 \\
HD 189733 b & 1.13 & 1.13 & 2.21 & 0.030 &  &K1-K2 & 0.80 & 0.75 & 7.67 & 2.97 &5050 & 41 , 42  &6.08$\pm$0.18 &10.3 &30.3 \\
HD 209458 b & 0.68 & 1.32 & 3.52 & 0.047 &  &G0 V & 1.01 & 1.14 & 7.65 & 3.75 &5942 & 43 , 44  &0.03$\pm$0.01 &10.5 &43.6 \\
HD 80606 b & 3.94 & 0.92 & 111. & 0.449 &  &G5 & 0.90 & 0.98 & 8.93 & 2.20 &5574 &45 &$<$1.80 &6.85 &35.2 \\
Kepler-4 b & 0.07 & 0.35 & 3.21 & 0.045 &  &G0 & 1.22 & 1.48 & 12.7 & 2.20 &5857 &46 & &9.40 &51.7 \\
Kepler-5 b & 2.11 & 1.43 & 3.54 & 0.050 &  & & 1.37 & 1.79 &  & 4.80 &6297 &47 & &10.1 & \\
Kepler-6 b & 0.66 & 1.32 & 3.23 & 0.045 &  & & 1.20 & 1.39 &  & 3.00 &5647 &48 & &10.4 & \\
Kepler-7 b & 0.43 & 1.47 & 4.88 & 0.062 &  &G0 & 1.34 & 1.84 &  & 4.20 &5933 &49 & &10.7 &61.7 \\
Kepler-8 b & 0.60 & 1.41 & 3.52 & 0.048 &  & & 1.21 & 1.48 & 13.9 & 10.5 &6213 &50 & &11.3 & \\
Kepler-9 b & 0.25 & 0.84 & 19.2 & 0.140 &  &G2 & 1.00 & 1.10 & 13.9 & 1.90 &5777 &51 & &8.89 &40.5 \\
Kepler-9 c & 0.17 & 0.82 & 38.9 & 0.225 &  &G2 & 1.00 & 1.10 & 13.9 & 1.90 &5777 &51 & &8.60 &40.5 \\
Kepler-9 d & 0.02 & 0.14 & 1.59 & 0.027 &  &G2 & 1.00 & 1.10 & 13.9 & 1.90 &5777 &90 & &9.15 &40.5 \\
Kepler-10 b & 0.01 & 0.12 & 0.83 & 0.016 &  &G & 0.89 & 1.05 & 10.9 & 0.50 &5627 &91 & &8.82 &37.8 \\
Lupus-TR-3 b & 0.81 & 0.89 & 3.91 & 0.046 &  &K1V & 0.87 & 0.82 & 17.4 &   &0.00 & &$<$0.64 & & \\
OGLE-TR-10 b & 0.63 & 1.26 & 3.10 & 0.041 &  &G or K & 1.18 & 1.16 &  & 7.70 &6220 &52 &0.08$\pm$0.04 &11.0 &47.0 \\
OGLE-TR-111 b & 0.53 & 1.06 & 4.01 & 0.047 &  &G or K & 0.82 & 0.85 &  & $<$5.00 &5070 &53 &$<$0.64 &10.5 &27.4 \\
OGLE-TR-113 b & 1.32 & 1.09 & 1.43 & 0.022 &  &K & 0.78 & 0.77 &  & $<$5.00 &4752 &52 &$<$1.80 &10.8 &28.1 \\
OGLE-TR-132 b & 1.14 & 1.18 & 1.68 & 0.030 &  &F & 1.26 & 1.34 &  & $<$5.00 &6411 &52 &$<$1.78 &10.7 &40.9 \\
OGLE-TR-182 b & 1.01 & 1.13 & 3.97 & 0.051 &  & & 1.14 & 1.14 & 16.8 &   &0.00 & &$<$1.76 & & \\
OGLE-TR-211 b & 1.03 & 1.36 & 3.67 & 0.051 &  & & 1.33 & 1.64 &  &   &0.00 & &$<$1.83 & & \\
OGLE-TR-56 b & 1.29 & 1.30 & 1.21 & 0.022 &  &G & 1.17 & 1.32 & 16.6 & 5.00 &5970 &52 &$<$1.79 &11.2 &48.7 \\
OGLE2-TR-L9 b & 4.50 & 1.61 & 2.48 & 0.030 &  &F3 & 1.52 & 1.53 &  & 39.3 &6933 &54 & &11.8 &50.6 \\
SWEEPS-04  & 3.80 & 0.81 & 4.20 & 0.055 &  & & 1.24 & 1.18 & 18.8 &   &0.00 & &$<$0.64 & & \\
SWEEPS-11  & 9.70 & 1.13 & 1.79 & 0.030 &  & & 1.10 & 1.45 & 19.8 &   &0.00 & &64.9$\pm$4.55 & & \\
TrES-1  & 0.61 & 1.08 & 3.03 & 0.039 &  &K0V & 0.87 & 0.83 & 11.7 & 1.08 &5214 &55 &$<$1.14 &9.71 &34.2 \\
TrES-2  & 1.20 & 1.17 & 2.47 & 0.035 &  &G0V & 0.98 & 0.94 & 11.4 & 2.00 &5850 &33 &$<$1.46 &9.98 &36.7 \\
TrES-3  & 1.91 & 1.32 & 1.30 & 0.022 &  &G & 0.92 & 0.81 & 12.4 & 1.50 &5650 &33 &$<$0.64 &10.2 &31.4 \\
TrES-4  & 0.91 & 1.79 & 3.55 & 0.050 &  &F & 1.38 & 1.84 & 11.5 & 8.50 &6200 &57 &$<$1.50 &11.3 &49.3 \\
WASP-1 b & 0.89 & 1.35 & 2.51 & 0.038 &  &F7V & 1.24 & 1.24 & 11.7 & $<$5.00 &6200 &58 &$<$0.64 &10.8 &36.7 \\
WASP-10 b & 3.06 & 1.08 & 3.09 & 0.037 &  &K5 & 0.71 & 0.78 & 12.7 & $<$6.00 &4675 &59 & &10.0 &27.8 \\
WASP-11 b & 0.46 & 1.04 & 3.72 & 0.043 &  &K3V & 0.82 & 0.81 & 11.8 & 0.50 &4980 &60 & &9.22 &31.4 \\
WASP-12 b & 1.41 & 1.79 & 1.09 & 0.022 &  &G0 & 1.35 & 1.57 & 11.6 & $<$2.20 &6300 &61 & &11.4 &59.9 \\
WASP-13 b & 0.46 & 1.21 & 4.35 & 0.052 &  &G1V & 0.00 & 1.34 & 10.4 & $<$4.90 &5826 &62 & &10.4 &47.5 \\
WASP-14 b & 7.72 & 1.25 & 2.24 & 0.037 &  &F5V & 1.31 & 1.29 & 9.75 & 4.90 &6475 &63 & &9.71 &40.5 \\
WASP-15 b & 0.54 & 1.42 & 3.75 & 0.049 &  &F5 & 1.18 & 1.47 & 10.9 & 4.00 &6300 &64 & &10.7 &42.8 \\
WASP-16 b & 0.85 & 1.00 & 3.11 & 0.042 &  &G3V & 1.02 & 0.94 & 11.3 & 3.00 &5700 &65 & &10.0 &35.5 \\
WASP-17 b & 0.49 & 1.74 & 3.73 & 0.051 &  &F6 & 1.20 & 1.38 & 11.6 & 9.00 &6550 &66 & &11.6 &43.1 \\
WASP-18 b & 10.4 & 1.16 & 0.94 & 0.020 &  &F6 & 1.28 & 1.21 & 9.30 & 11.0 &6400 &67 & &10.5 &37.9 \\
WASP-19 b & 1.15 & 1.31 & 0.78 & 0.016 &  &G8V & 0.95 & 0.93 & 12.3 & 4.00 &5500 &68 & &11.7 &33.2 \\
WASP-2 b & 0.91 & 1.01 & 2.15 & 0.031 &  &K1V & 0.84 & 0.78 & 11.9 & $<$5.00 &5200 &58 &$<$0.64 &10.5 &32.5 \\
WASP-21 b & 0.30 & 1.07 & 4.32 & 0.052 &  &G3V & 1.01 & 1.06 & 11.6 & 1.50 &5800 &69 & &9.97 &39.7 \\
WASP-22 b & 0.56 & 1.12 & 3.53 & 0.046 &  &G & 1.10 & 1.13 & 12.0 & 3.50 &6000 &70 & &10.3 &43.7 \\
WASP-24 b & 1.03 & 1.10 & 2.34 & 0.035 &  &F8-9 & 1.12 & 1.14 & 11.3 & 6.96 &6075 &71 & &10.7 &33.6 \\
WASP-25 b & 0.58 & 1.26 & 3.76 & 0.047 &  &G4 & 1.00 & 0.95 & 11.9 & 3.00 &5750 &72 & &10.4 &36.1 \\
WASP-26 b & 1.02 & 1.32 & 2.75 & 0.040 &  &G0 & 1.12 & 1.34 & 11.3 & 2.40 &5950 &73 & &10.2 &49.0 \\
WASP-28 b & 0.91 & 1.12 & 3.40 & 0.045 &  &F8-G0 & 1.08 & 1.05 & 12.0 & 2.20 &6100 &64 & &9.86 &31.6 \\
WASP-29 b & 0.24 & 0.74 & 3.92 & 0.045 &  &K4V & 0.82 & 0.84 & 11.3 & 1.50 &4800 &74 & &9.64 &30.6 \\
WASP-3 b & 1.76 & 1.39 & 1.84 & 0.031 &  &F7V & 1.24 & 1.35 & 10.6 & 13.4 &6400 &33 &$<$0.64 &11.3 &41.0 \\
WASP-33 b & 4.10 & 1.49 & 1.21 & 0.025 &  &A5 & 1.49 & 1.44 & 8.30 & 90.0 &7430 &76 & &12.4 &28.5 \\
WASP-37 b & 1.69 & 1.13 & 3.57 & 0.043 &  &G2 & 0.84 & 0.97 & 12.7 & 2.40 &5800 &77 & &9.67 &37.3 \\
WASP-38 b & 2.71 & 1.07 & 6.87 & 0.075 &  &F8 & 1.21 & 1.36 & 9.42 & 8.60 &6150 &78 & &9.64 &38.9 \\
WASP-4 b & 1.12 & 1.41 & 1.33 & 0.023 &  &G7V & 0.90 & 1.15 & 12.6 & 2.20 &5500 &79 &$<$0.64 &10.9 &38.9 \\
WASP-5 b & 1.63 & 1.17 & 1.62 & 0.027 &  &G4V & 1.02 & 1.08 & 12.2 & 3.40 &5700 &80 &$<$0.64 &10.4 &39.3 \\
WASP-6 b & 0.50 & 1.22 & 3.36 & 0.042 &  &G8 & 0.00 & 0.87 & 12.4 & 1.40 &5450 &81 & &9.89 &31.2 \\
WASP-7 b & 0.96 & 0.91 & 4.95 & 0.061 &  &F5V & 1.28 & 1.23 & 9.51 & 17.2 &6400 &82 & &10.5 &38.4 \\
WASP-8 b & 2.23 & 1.17 & 8.15 & 0.079 &  &G6 & 1.03 & 0.95 & 9.90 & 2.00 &5600 &83 & &8.90 &34.8 \\
XO-1 b & 0.90 & 1.18 & 3.94 & 0.048 &  &G1V & 1.00 & 0.92 & 11.3 & $<$3.00 &5750 &84 &$<$0.64 &10.0 &35.4 \\
XO-2 b & 0.57 & 0.97 & 2.61 & 0.036 &  &K0V & 0.98 & 0.96 & 11.1 & $<$2.30 &5340 &85 &$<$0.64 &10.1 &39.7 \\
XO-3 b & 11.7 & 1.21 & 3.19 & 0.045 &  &F5V & 1.21 & 1.37 & 9.80 & 18.5 &6429 & 86 , 87  &$<$0.64 &10.0 &41.9 \\
XO-4 b & 1.72 & 1.34 & 4.12 & 0.055 &  &F5V & 1.32 & 1.55 & 10.7 & 8.80 &6397 &88 & &10.4 &45.4 \\
XO-5 b & 1.07 & 1.08 & 4.18 & 0.048 &  &G8V & 0.88 & 1.06 & 12.1 & 0.70 &5370 &89 & &8.96 &35.3 \\
\hline
\multicolumn{15}{l}{\parbox{22cm}{\textbf{References.} (1) Flux from Kashyap \etal\ 2008 in units of $10^{-13}$~erg~s$^{-1}$~cm$^{-2}$; (2) Barge \etal\ 2008; (3) Bonomo \etal\ 2010; (4) Gandolfi \etal\ 2010; (5) Gillon \etal\ 2010; (6) Cabrera \etal\ 2010; (7) Bouchy \etal\ 2008; (8) Deleuil \etal\ 2008; (9) Moutou \etal\ 2008; (10) Rauer \etal\ 2009; (11) Fridlund \etal\ 2010; (12) L\'eger \etal\ 2009; (13) Bord\'e \etal\ 2010; (14) Deeg \etal\ 2010; (15) Charbonneau \etal\ 2009; (16) Butler \etal\ 2004; (17) Torres 2007; (18) Bakos \etal\ 2007a; (19) Bakos \etal\ 2010a; (20) Hartman \etal\ 2009; (21) Bakos \etal\ 2009b; (22) Torres \etal\ 2010; (23) Kov\'acs \etal\ 2010; (24) Buchhave \etal\ 2010; (25) Howard \etal\ 2010; (26) Hartman \etal\ 2010a; (27) Bakos \etal\ 2007b; (28) Bakos \etal\ 2010b; (29) Kipping \etal\ 2010; (30) Quinn \etal\ 2010; (31) Hartman \etal\ 2010b; (32) Torres \etal\ 2007; (33) Christiansen \etal\ 2010; (34) Bakos \etal\ 2007c; (35) Noyes \etal\ 2008; (36) P\'al \etal\ 2008; (37) Latham \etal\ 2009; (38) Shporer \etal\ 2009; (39) Sato \etal\ 2005; (40) Winn \etal\ 2009; (41) Bouchy \etal\ 2005; (42) Winn \etal\ 2007; (43) Queloz \etal\ 2000; (44) Brown \etal\ 2001; (45) Moutou \etal\ 2009; (46) Borucki \etal\ 2010; (47) Koch \etal\ 2010; (48) Dunham \etal\ 2010; (49) Latham \etal\ 2010; (50) Jenkins \etal\ 2010; (51) Holman \etal\ 2010; (52) Bouchy \etal\ 2004; (53) Pont \etal\ 2004; (54) Snellen \etal\ 2009; (55) Laughlin \etal\ 2005; (57) Sozzetti \etal\ 2009; (58) Collier-Cameron \etal\ 2007; (59) Christian \etal\ 2009; (60) Bakos \etal\ 2009a; (61) Hebb \etal\ 2009; (62) Skillen \etal\ 2009; (63) Joshi \etal\ 2008; (64) West \etal\ 2009; (65) Lister \etal\ 2009; (66) Anderson \etal\ 2010; (67) Hellier \etal\ 2009a; (68) Hebb \etal\ 2010; (69) Bouchy \etal\ 2010; (70) Maxted \etal\ 2010; (71) Street \etal\ 2010; (72) Enoch \etal\ 2010; (73) Smalley \etal\ 2010; (74) Hellier \etal\ 2010; (76) Collier-Cameron \etal\ 2010; (77) Simpson \etal\ 2011; (78) Barros \etal\ 2011; (79) Wilson \etal\ 2008; (80) Anderson \etal\ 2008; (81) Gillon \etal\ 2009; (82) Hellier \etal\ 2009b; (83) Queloz \etal\ 2010; (84) McCullough \etal\ 2006; (85) Burke \etal\ 2007; (86) Johns-Krull \etal\ 2008; (87) H\'ebrard \etal\ 2008; (88) McCullough \etal\ 2008; (89) P\'al \etal\ 2009; (90) Torres \etal\ 2011; (91) Batalha \etal\ 2011.}}
\end{longtable}
\end{landscape}
}

\end{document}